# The Ferroelectric-Gate Fin Microwave Acoustic Signal Processor

Faysal Hakim, Nicholas Rudawski, Troy Tharpe, Roozbeh Tabrizian[*]

*Department of Electrical and Computer Engineering, Herbert Wertheim College of Engineering, University of Florida, Gainesville, FL 32611, USA*

## Abstract

Wireless communication through dynamic spectrum allocation over microwave bands, essential to accommodate exponentially growing data traffic, requires massive array of radio-frequency (RF) filters for adaptive signal shaping at arbitrary frequencies. However, conventional RF filters based on planar acoustic resonators are incapable to realize such massive integrated arrays, due to their large footprint and limited on-chip frequency scalability. Here, we present a signal processor enabled by integration of three-dimensional ferroelectric-gate fin (FGF) nano-acoustic resonators with extreme frequency tailorability and large-scale integrability. FGFs are created by growing atomic-layered ferroelectric hafnia-zirconia transducers on silicon nano-fins, operate in bulk acoustic modes with lithographically scalable frequency over 3-28 GHz, and provide record-high frequency – quality factor – electromechanical coupling product ($f \times Q \times k_t^2$) of $19.4 \times 10^{10}$ (at ~11GHz). A monolithic filter-array covering 9-12 GHz is also demonstrated by on-chip electrical coupling of FGFs. This demonstration highlights the potential of FGF resonators to realize chip-scale adaptive processors extendable to millimeter-wave frequencies.

[*]Corresponding Author.  Email: rtabrizian@ece.ufl.edu

## Main

Wireless systems rely on numerous preset bands in radio-frequency (RF) spectrum to simultaneously communicate exabytes of data for different applications and among billions of users, with minimum latency and interference[1-3]. The spectrum management of such massive traffic is enabled by a large set of static RF filters, corresponding to preset communication bands, that are orchestrated through a set of switches at the front-end of wireless systems[4-6]. Currently, mobile systems rely on more than a hundred of static filters for RF spectral processing over 0.3 GHz – 7 GHz[7-9]. These filters are created from electrical coupling of surface and bulk acoustic wave (SAW and BAW) resonators, to benefit from their small size compared to electrical and optical counterparts, which is due to the significantly lower acoustic wave velocity compared to electromagnetic waves. SAW and BAW filters with different frequencies and bandwidths are packaged in various chips and heterogeneously integrated with RF switches to form multi-band signal processors in RF front-end circuitry.

However, the current signal processing architecture based on static filtering is not scalable with exponential increase in wireless users, due to emergence of the internet of things, as well as progressive demand for higher data rates and bandwidths in new applications such as the metaverse. To accommodate this ever-growing data traffic, new wireless generations target the use of dynamic spectrum allocation, instead of relying on static bands, to substantially enhance the spectrum-use efficiency[10-12]. Dynamic spectrum allocation requires highly adaptive signal processors that provide extensive configurability of filter type, frequency, and bandwidth, over a large microwave spectrum stretching to the millimeter-wave regime.

A widely adaptive microwave signal processor requires a resonator technology that either provides extensive tunability, to cover a wide spectrum of interest using one or few elements, or enables massive arraying and extreme frequency scalability in a reasonable integrated footprint[13-15]. This cannot be implemented using conventional piezoelectric SAW and BAW resonators, since they suffer from at least two of the following deficiencies. First, they do not provide the frequency tunability required for wideband adaptation. Second, they offer limited frequency tailorability within fabrication process (*e.g.*, through lithographical scaling of geometry) when implemented on the same chip. Third, their planar architecture inherently limits massive arraying as it imposes excessive footprint consumption and makes large-scale integration challenging.

In this work, we demonstrate the use of the third geometrical dimension to create non-planar acoustic resonators that surpass fundamental limits of planar counterparts for massive arraying and frequency scaling. Similar to non-planar transistors (*i.e.*, fin field effect transistors), extending the functional volume of acoustic resonators to the third dimension enables substantial enhancement in integration density without reducing interface area (*i.e.*, electromechanical transduction area). Further, the three-dimensional resonator structure benefits from new acoustic resonance modes that offer lithographical frequency tailorability independent from piezoelectric film thickness, which is constant across the wafer and limits scaling of conventional BAW resonators[16, 17].

The presented resonators are based on high-aspect-ratio silicon (Si) nano-fins that are wrapped around by an atomic-layered hafnia-zirconia ($Hf_{0.5}Zr_{0.5}O_2$, or simply HZO) ferroelectric-gate that serves for piezoelectric transduction. The resulting ferroelectric-gate fin (FGF) architecture enables excitation of lateral BAW modes with a frequency defined lithographically by nano-fin width; hence, enabling dense integration of resonator arrays covering a wide spectrum, on a single chip. The FGF nano-acoustic resonators serve as building blocks to provide poles and zeros at arbitrary frequencies, thus facilitating realization of highly adaptive microwave signal processors for modern wideband dynamic wireless systems. In this paper, FGF nano-acoustic resonators over 3-27 GHz with frequency-quality factors ($f{\times}Q$) as high as $0.76{\times}10^{13}$ and electromechanical coupling – quality factor product ($k_t^2{\times}Q$) as large as 17.7 are demonstrated. FGF resonators with different fin widths are electrically coupled on the chip to create bandpass filter arrays with arbitrary center-frequency and bandwidth over 9-12 GHz, highlighting capability of FGFs to create chip-scale adaptive microwave signal processors for dynamic spectrum use.

## FGF Nano-Acoustic Resonator Concept and Scaling

FGF nano-acoustic resonators are created by conformal covering of a high-aspect-ratio semiconductor rectangular parallelopiped with a piezoelectric transducer. Such a structure enables excitation of various harmonics of width-extensional BAW modes (*i.e.*, WE$_n$, $n \in \mathbb{N}$), created from propagation of planar waves in the width direction and reflection at the stress-free surfaces on the two sides. The conformal transducer is composed of a piezoelectric film sandwiched between two metal electrodes that enable application of electric-field across for electromechanical excitation of acoustic waves and mechanical resonance modes. To achieve efficient transduction of WE$_n$ modes, the piezoelectric film is required to have a polar axis perpendicular to the sidewall of fin. Further, when the polar axis of transducers on both sidewalls are aligned either inward or outward, the excitable modes are limited to odd harmonics (*i.e.*, $n = 2k + 1, k \in \mathbb{Z}$).

FGF resonators benefit from a large transduction area that is scalable by fin height as well as number of fins that are closely packed and electrically coupled. Figure 1 (a) shows the schematic of a one-port FGF resonator composed of three parallel fins with the same dimensions. In this architecture, two ferroelectric-gates (*i.e.*, RF terminals) are placed across the fins and electrically coupled through floating bottom electrodes covering each fin. This allows extension of the electric field across piezoelectric film thickness to achieve higher transduction efficiency. Figure 1 (a) inset shows the proposed symbol for the one-port FGF resonator created from three fins. Figure 1 (b) shows the cross-sectional mode-shapes of $WE_{1,3,5}$ modes in each fin, simulated using COMSOL for a resonator with Si fin width ($W_{fin}$) of 500nm, HZO piezoelectric film thickness ($T_{HZO}$) of 100nm, and tungsten (W) electrode thickness of 50nm. As evident in the mode shapes, the energy of $WE_n$ modes are localized in the fin volume, with minimum leakage to the supporting substrate. Further, except for the top and bottom of the fin, the mechanical stress field is uniform over the entire height of the fin, enabling efficient piezoelectric coupling of BAW modes with minimum charge cancellation.

The frequency of $WE_n$ modes (*f*) in FGF resonators is defined by the path of BAW as propagating across the resonator width, consisting of semiconductor fin width as well as the transducer thickness. Therefore, *f* can be lithographically scaled by changing the fin width. Considering the variation of BAW velocity in different materials across resonator width, the frequency scaling with $W_{fin}$ does not follow a simple closed-form expression. In this work, Mason's waveguide modeling approach is used to predict performance metrics of $WE_n$ modes for different $W_{fin}$ and $T_{HZO}$, and based on fundamental material properties. In this model, each constituent layer across resonator width is represented by frequency-dependent lumped electrical impedances that represent equivalent mass, stiffness, and damping as well as electromechanical transduction (in piezoelectric layers). This simplifies the FGF nano-acoustic resonator to an equivalent circuit model, enabling prediction of resonator admittance, and extraction of frequency, *Q*, and $k_t^2$ of all $WE_n$ harmonics (see Supplementary Information, section S1, for detailed discussion on the modeling approach).

Figure 1 (c) shows the scaling plots for WE$_3$ mode, as a function of W$_{fin}$ and for different T$_{HZO}$. In frequency scaling plot, two regions with different characteristics are identified. In the first region where fin-width is significantly larger compared to transducer thickness, the frequency is scaled inversely linear with W$_{fin}$. In the second region corresponding to narrower fins and higher frequencies, frequency scaling deviates from inversely linear relation. This corresponds to an increased effect of acoustic-impedance mismatch between Si and HZO with W at small BAW wavelengths, resulting in large reflection of the wave at material boundaries, which translates to a smaller effective paths and higher frequencies. Further, it is evident that extreme frequency scaling to mm-wave regime requires reduction in HZO thickness to 20nm. This is perfectly feasible considering extreme thickness scalability of HZO films to few nanometers while sustaining a large polarization and piezoelectric coupling[18, 19]. Figure 1 (c) also shows the scaling characteristics for $k_t^2$ and $Q$ in nano-fin resonators with different W$_{fin}$ and T$_{HZO}$, highlighting an opposite trend. $k_t^2$ increases with relative reduction of W$_{fin}$ to T$_{HZO}$, as a result of relative increase in the volume of electromechanically active (*i.e.*, piezoelectric) material in resonator structure. On the other hand, resonator $Q$ reduces with relative reduction of W$_{fin}$ to T$_{HZO}$, corresponding to the higher acoustic energy dissipation rate in polycrystalline HZO film compared to single-crystal Si fin[20-22].

**FGF Nano-Acoustic Resonator Implementation and Characterization**

The performance of an FGF resonator correlates closely with its three-dimensional geometry as well as mechanical, electrical, and electromechanical properties of constituent materials. First, efficient excitation of BAW modes with uniform energy distribution and minimum charge cancellation, essential for linear operation and high $k_t^2$, requires a constant fin width and transducer thickness across the entire height. This becomes excessively challenging when increasing the aspect-ratio of fins to enhance their transduction area and reduce resonator impedance. Second, the interfaces of constituent layers in nano-fin resonator should be as smooth as possible to minimize surface scattering of BAW that degrades resonator $Q$[23-25]. This is especially crucial when targeting an increase in frequency that corresponds to reduced BAW wavelength and higher surface scattering. Third, efficient electromechanical transduction for high $k_t^2$ requires accurate control over the crystallinity and polarity of piezoelectric film grown on the sidewall. This is highly challenging as conventional piezoelectric film-growth techniques are optimized for high quality planar films[26, 27]. In these techniques, films grown on non-planar surfaces have poor texture and crystallinity, or polar-axis misorientation from surface normal. Forth, as an inherently three-dimensional device, creation of FGF resonators require dimensionally selective patterning of metal electrodes to form well-defined RF terminals with reduced static electrical feedthrough.

In this work, a fabrication process is developed to realize stringent geometrical and material requirements, as well as three-dimensional electrode patterning, to create high-performance FGF nano-acoustic resonators and filters. The fabrication process consists of three key steps. First, Si fins with ultra-high aspect-ratios are created by chemical etching with crystallographic selectivity. The fins are created on (110) silicon-on-insulator substrates, and are aligned to [112] crystallographic direction. When exposed to basic wet etchants (*e.g.*, potassium hydroxide, tetramethylammonium hydroxide), such crystallographic alignment enables formation of rectangular parallelopipeds with perfectly straight sidewalls, due to the near-perfect etch stop at [111] planes. This approach not only enables creation of ultra-high aspect-ratio fins with constant width across the height, but also yield ultra-smooth Si sidewall surfaces.

Second, the FGF nano- acoustic resonators of this work benefit from piezoelectric transducer that is entirely grown using conformal atomic layer deposition (ALD). The centerpiece of transducer, *i.e.*, the ferroelectric film with large piezoelectricity, is created from formation of HZO superlattice created by periodic stacking with alumina ($Al_2O_3$) interlayers. HZO superlattice is recently demonstrated as a hence piezoelectric transducer with large electromechanical transduction, high acoustic velocity, and low energy dissipation[20, 28-30]. The conformal ALD growth enables achieving ferroelectric-gate transducers with uniform thickness, texture, crystallinity, and polarization over the entire sidewall of Si fins. Further, atomic-layered ferroelectric HZO film have shown exceptional preservation of their polarization and piezoelectricity even when scaled to few nanometers. These make HZO superlattice an ideal choice for creation of FGF nano- acoustic resonators with extreme frequency scalability to mm-wave regime. In this work, HZO superlattice with thickness of 100nm (*i.e.*, ten 9nm-thick HZO layers laminated with nine 1nm-thick $Al_2O_3$ interlayers) is sandwiched between 50nm-thick W electrodes that are also deposited using ALD.

Finally, the three-dimensional bottom and top W electrode are patterned using anisotropic and isotropic dry etching, respectively. High dimensional selectively is achieved by using photoresist masks with large aspect-ratio. The detailed information on chemicals and techniques used for fabrication of FGF nano-acoustic resonators and filters are presented in Supplementary Information, section S2.

Figure 2 (a, b, c) shows the scanning electron microscope (SEM) images of FGF resonators formed from a single or multiple fins, from different angles. For the multi-fin structure, the continuity of top electrodes (*i.e.*, three-dimensional gates) over the five fins is captured by top-view SEM image. Also, the three-dimensional profile of the top electrode is evident in side-view SEM images. Figure 2 (c, d) shows cross-sectional SEM and transmission electron microscope (TEM) images of FGF resonators cross-section, for devices with different Si fin widths of 186nm and 707nm, and fin height of ~6300nm. The very large aspect ratios, as high as 34:1, while having a perfectly straight and smooth Si sidewall, is evident in images. Figure 2 (c, d) also shows images zoomed on top and bottom corners of the resonators. These images show the perfectly isolated bottom W electrodes that only exist on the sidewalls, the conformal HZO superlattice transducer, and the conform top W electrodes. Figure 2 (e) shows the TEM image zoomed on HZO superlattice, highlighting the constituent layers as well as polar orthorhombic texture in grains of two different HZO layers.

FGF resonators with Si fin width ranging from 300nm to 1000nm are implemented and their performance is characterized using RF electrical measurements. Prior to RF characterization, HZO transducer is poled, through application of 1kHz bipolar pulses, to enhance linear transduction with large longitudinal piezoelectric coefficient (*i.e.*, $d_{33}$)[30, 31]. See Supplementary Information, section S3, for detailed discussion on transducer poling. Following transducer poling, one-port FGF resonators are characterized by measuring their reflection coefficient (*i.e.*, $S_{11}$), when excited with -5dBm input power. Resonator admittance is extracted from $S_{11}$ response and performance metrics of BAW modes are measured. Figure 3 (a, b) shows the measured admittance of FGF resonators with 300nm and 1000nm Si fin widths, over a wide frequency span. The measured responses are compared with loss-less Mason's model, to enable identification of BAW mode corresponding to each resonance peak. For the resonator with 1000nm Si fin width, $WE_3$ and $WE_5$ modes are evident at 8.55 GHz and 11.86 GHz, respectively. For the resonator with 300nm Si fin width, $WE_1$, $WE_3$, and $WE_5$ modes are evident at 3.07 GHz, 11.61 GHz, and 27.29 GHz, respectively. A good match between measurement and model is evident for both fin, which is a result of well-defined fin geometry and accurate control over transducer thickness and acoustic properties. Figure 3 (c) shows the measured admittance of FGF resonators with different Si fins covering the 300nm to 1000nm range, all implemented in the same batch, highlighting evident frequency scaling of $WE_3$ BAW mode with fin widths to cover the entire 9 GHz to 12 GHz. Figure 3 (d) shows the measured admittance of selective peaks over 3-30 GHz for FGF resonators with different fin widths and operating in different WE modes.

FGF resonators show $Q$s as high as 1148 (for $W_{fin}$ of 300mn operating in $WE_1$ mode) and $k_t^2$s as high as 4.94% (for $W_{fin}$ of 300nm operating in $WE_3$ mode). The highest frequency peaks, over 18 GHz – 28 GHz corresponds to $WE_5$ modes in FGF resonator with fin widths over 300nm – 500nm, showing $Q$s over 77 – 404 and $k_t^2$s over 0.27% – 0.65%. The highest frequency-$Q$ product ($f \times Q$) of $0.76 \times 10^{13}$ is measured for the resonator with 500nm fin width at 18.78 GHz ($WE_5$ mode). The highest $k_t^2 \times Q$ of 17.73 is measured for the resonator with 500nm fin width at 10.96 GHz ($WE_3$ mode). See Supplementary Information, section S3, for detailed performance merits of WE modes for resonators with different fin widths.

## FGF Nano-Acoustic Filter Array

FGF resonators are electrically coupled to form bandpass filters. Resonators with different fin width are connected through planar routings to form 2.5-stage ladder filters. An array of filters is implemented with lithographically tailorable center-frequency and bandwidth, to realize a signal processor for dynamic spectrum control over 9-12 GHz. Figure 4 shows SEM images of an FGF filter from different angles. The filter is created from three series resonators and two shunt resonators, electrically connected to form the 2.5-stage ladder. The frequency and transduction area of series and shunt resonators in the ladder are optimized for proper passband impedance matching and isolation, needed for high-selectivity shape-factors[32]. This include optimizing Si fin widths to overlap the parallel resonance of shunt resonator with series resonance of series resonator in ladder filter. The transduction area of shunt resonators are also optimized to be nearly twice that of series, by changing corresponding gate widths, to ensure large out of band rejection.

Figure 5 shows the measured transmission responses magnitude ($|S_{21}|$) for FGF filters with different frequencies over 9-12 GHz, created from resonators with different fin widths. The change in filter bandwidth at different center-frequencies corresponds to the change in $k_t^2$ with the scaling of Si fin width. This demonstration highlights the potential of FGF technology to create single-chip configurable wideband signal processors. The detailed performance merits of filters in fig. 5 are presented in Supporting Information, section S5.

In all the filters, large ripples are evident over the passband, as a result of spurious modes in resonator admittances that correspond to Lamb waves propagating in height or length direction of the fin. These spurs may be suppressed by proper dispersion engineering through opting for geometrical variations in Si fin across its height and length[33]. Further, the large insertion-loss of filters can be reduced by opting for multi-fin resonator architecture, to scale the admittances for proper matching with 50Ω electrical terminations, as well as further densification of arraying to reduce excessive loss due to planar routing lines.

## Conclusion

We have reported a three-dimensional nano-acoustic resonator technology that enables large-scale integration of RF filters with lithographically defined performance, to enable realization of single-chip dynamic signal processors with extreme frequency scalability. The resonators are created by covering high aspect-ratio Si fins with atomic-layered ferroelectric HZO transducer gates. These ferroelectric-gate fin (FGF) nano-acoustic resonators operate in lateral BAW modes with frequency, $Q$, and $k_t^2$ scalable lithographically by Si fin width as well as transducer thickness over cm- and mm-wave regimes. A waveguide-based analysis is presented to accurately predict scaling characteristic of FGF resonators and enable proper filter design. The fabrication process developed for creation of FGF resonators, consisting of crystallographic-dependent chemical etching and atomic layer deposition techniques to realize fin structures with ultra-high aspect-ratio and straight and smooth sidewalls, and conformal piezoelectric transducers with uniform thickness and acoustic properties over the fin sidewalls. FGF resonators with frequencies over 3-27 GHz are demonstrated with $f \times Q$s as high as $0.76 \times 10^{13}$ and $k_t^2 \times Q$s as high as 17.73. FGF resonators different fin-widths are electrically coupled to form bandpass filter arrays with lithographically defined center-frequency and bandwidth, covering 9-12 GHz. The presented microwave acoustic resonator technology shows a promising perspective for realization of a chip-scale widely adaptive signal processors for dynamic spectrum allocation in future wireless systems.

## Methods

**Wafer specification:** For fabrication of nano fins, silicon-on-insulator (SOI) wafers are used. For crystallographic orientation dependent etching of fins, a 6μm-thick (110)-oriented device layer with <110> orientation and <112> primary flats is used. The buried oxide (Box) and handle layer thicknesses are 500nm and 500μm respectively. The handle layer had a resistivity exceeding 10,000 ohm.cm to reduce the effect of substrate feedthrough.

**Film deposition:**

ALD of $Hf_{0.5}Zr_{0.5}O_2$ – $Al_2O_3$ superlattice: Amorphous $Hf_{0.5}Zr_{0.5}O_2$ - $Al_2O_3$ superlattice transducers are deposited using a Cambridge NanoTech Fiji 200 atomic layer deposition system with a process setpoint of 200°C. 9.2nm-thick layers of $Hf_{0.5}Zr_{0.5}O_2$ are deposited using 56 cycles of tetrakis(dimethylamido)hafnium(IV) (TDMAH) and tetrakis(dimethylamido)zirconium(IV) (TDMAZ) precursors pulsed with a 1:1 ratio. 300W hydrogen and oxygen plasmas are applied following each monolayer deposition for precursor oxidation with enhanced orthorhombic phase concentration[34]. Next, 1nm-thick $Al_2O_3$ is thermally deposited using 10 cycles of trimethylaluminum (TMA) precursor to limit $Hf_{0.5}Zr_{0.5}O_2$ vertical grain size to sub-10nm, during further deposition steps. This 9.2nm $Hf_{0.5}Zr_{0.5}O_2$ – 1nm $Al_2O_3$ deposition sub-cycle is then repeated additional times (depending on targeted thickness), followed by a final 9.2nm deposition of $Hf_{0.5}Zr_{0.5}O_2$.

ALD of W electrode: Amorphous W electrodes are deposited in ALD chamber using silane ($SiH_4$) and tungsten hexafluoride ($WF_6$) gas.

**Device fabrication:** Supplementary Fig. S2 depicts the three-mask fabrication process flow used to implement nano-fin acoustic resonators. First, 200nm silicon oxide ($SiO_2$) is patterned on SOI wafer via optical lithography to serve as hard mask for fin etching. Then, the sample is submerged in 2.3% tetra methyl ammonium hydroxide (TMAH) solution at 65°C for 15 mins to etch 6.5μm silicon and define fins. This is followed by removal of $SiO_2$ mask in buffered oxide etchant. Next, a 50nm-thick W is deposited using ALD. The deposited W is anisotropically etched in reactive ion etching (RIE) chamber using sulfur hexafluoride ($SF_6$) and argon (Ar) gas mixture to remove W layer on all planar surfaces, while keeping it on sidewalls of fins. After RIE of W to serve as floating bottom electrode in FGF, a 100nm-thick $Hf_{0.5}Zr_{0.5}O_2$ – $Al_2O_3$ superlattice is deposited by ALD, followed by another 50nm-thick W layer. Next, the wafer is subjected to rapid thermal annealing (RTA) using SSI Solaris 150 tool for 20s at 500°C in $N_2$ ambient, to promote preferred morphology in the superlattice film (*i.e.*, polar orthorhombic phase). Then, the top W layer is patterned isotropically in RIE chamber, by $SF_6$ gas, using 10μm-thick photoresist mask. Finally, 500nm-thick gold (Au) layer is deposited by lift-off to serve for routings and probing pads.

**Device characterization:**

DC analysis: Polarization – electric field (P-E) hysteresis loop of W/$Hf_{0.5}Zr_{0.5}O_2$ – $Al_2O_3$/W sidewall capacitors with different areas are obtained using Radiant PiezoMEMS Analyzer. For this analysis, a 45V, 1 kHz square pulse is applied between the two ports of resonator for 1s to wake the film up. Then, a 70V bipolar (triangular) pulse train at 10 kHz is applied to obtain the P-E loop.

RF measurement and analysis: After cycling and setting DC operating point, the FGF resonators are measured using a Keysight N5222A vector network analyzer with an input power -5 dBm.

**SEM imaging:** SEM images are taken using FEI Nova NanoSEM 430 tool.

**S/TEM imaging:**

Sample Preparation: A modified in-situ lift-out process using an FEI Helios G4 CXe dual beam plasma focus ion beam/scanning electron microscope equipped with a Pt GIS and EasyLift in-situ micromanipulator is used to prepare samples for S/TEM imaging. Electron-beam induced deposition of Pt is used to produce a ~2 μm-thick conformal protective coating on the whole surface of the device. The grid is then in-situ lifted out and welded to a Ted Pella Mo lift-out grid. After attaching to the grid, the grid (remaining in its grid holder) is removed from the stage and then manually rotated up to 45 degrees by hand in its grid holder and then reloaded onto the stage; this results in the tall direction of the fin being rotated away from the FIB during subsequent final milling (by the same amount it is rotated in the grid holder), which is necessary to produce a lamella with the most uniform thickness and minimal curtaining to allow accurate inspection of the film quality at all points on the fin surface. Thinning is initially started at 30 kV and then completed at 8 kV to reduce damage layer thickness.

Sample Imaging: An FEI Themis Z scanning/transmission electron microscope (S/TEM) with Cs probe correction, equipped with a Fischione Instruments Model 3000 high-angle annular dark-field (HAADF) STEM detector and FEI Ceta 16 megapixel CMOS camera is used to perform HAADF-STEM and high-resolution (lattice) imaging (HR-TEM).

## Data Availability

The authors affirm that the main data supporting the findings of this article are available within the manuscript and its Supplementary Information. Extra data are available from the corresponding author upon request.

## Acknowledgement

The authors would like to thank Dr. Timothy Hancock for technical discussions and support of this effort, and the University of Florida Nanoscale Research Facility cleanroom staff for fabrication support. F.H., T.T., and R.T. acknowledge the financial support from the Defense Advanced Research Projects Agency (DARPA) through the Young Faculty Award (Grant D19AP00044) and National Science Foundation (NSF) through the CAREER award (Grant ECCS-1752206).

# References

1   S. Dang, O. Amin, B. Shihada, M. S. Alouini, What should 6G be?, *Nature Electronics* **3**, 20-29 (2020).

2   S. Borkar, H. Pande, Application of 5G next generation network to Internet of Things, *International Conference on Internet of Things and Applications (IOTA)* **2016**, 443-447.

3   Y. Wang, J. Li, L. Huang, Y. Jing, A. Georgakopoulos, P. Demestichas, 5G Mobile: Spectrum Broadening to Higher-Frequency Bands to Support High Data Rates, *IEEE Vehicular Technology Magazine*, **9**, 39-46 (2014).

4   A. O. Watanabe, M. Ali, S. Y. B. Sayeed, R. R. Tummala, M. R. Pulugurtha, A Review of 5G Front-End Systems Package Integration. *IEEE Transactions on Components, Packaging and Manufacturing Technology*, **11**, 118-133 (2021).

5   S. Mahon, The 5G Effect on RF Filter Technologies. *IEEE Transactions on Semiconductor Manufacturing*, **30**, 494-499 (2017).

6   R. Ruby, A Snapshot in Time: The Future in Filters for Cell Phones, *IEEE Microwave Magazine*, **16**, 46-59 (2015).

7   R. Aigner, G. Fattinger, 3G – 4G – 5G: How Baw Filter Technology Enables a Connected World. *2019 20th International Conference on Solid-State Sensors, Actuators and Microsystems & Eurosensors XXXIII (TRANSDUCERS & EUROSENSORS XXXIII)* **2019**, 523-526.

8   R. Aigner, G. Fattinger, M. Schaefer, K. Karnati, R. Rothemund, F. Dumont, BAW Filters for 5G Bands. *2018 IEEE International Electron Devices Meeting (IEDM)* **2018**, 14.5.1-14.5.4.

9   D. Kim, G. Moreno, F. Bi, M. Winters, R. Houlden, D. Aichele, J. B. Shealy, Wideband 6 GHz RF Filters for Wi-Fi 6E Using a Unique BAW Process and Highly Sc-doped AlN Thin Film, *2021 IEEE MTT-S International Microwave Symposium (IMS)* **2021,** 207-209.

10  W. S. H. M. W. Ahmad, N. A. M. Radzi, F. S. Samidi, A. Ismail, F. Abdullah, M. Z. Jamaludin, M. Zakaria, 5G technology: Towards dynamic spectrum sharing using cognitive radio networks. *IEEE Access*, **8**, 14460-14488 (2020).

11  J. Zhu, K. J. R. Liu, Cognitive radios for dynamic spectrum access-dynamic spectrum sharing: A game theoretical overview, *IEEE Communications Magazine*, **45.5***,* 88-94 (2007).

12  CL. I, S. Han, S. Bian, Energy-efficient 5G for a greener future, *Nature Electronics*, **3**, 182–184 (2020).

13  R. Gómez-García, A. C. Guyette, Reconfigurable Multi-Band Microwave Filters, *IEEE Transactions on Microwave Theory and Techniques*, **63**, 1294-1307 (2015).

14  D. Marpaung, J. Yao, J. Capmany, Integrated microwave photonics. *Nature Photonics,* **13**, 80–90 (2019).

15  J. Fandiño, P. Muñoz, D. Doménech, J. Capmany, A monolithic integrated photonic microwave filter. *Nature Photonics*, **11**, 124–129 (2017).

16  J. Anderson, Y. He, B. Bahr, D. Weinstein, Integrated acoustic resonators in commercial fin field-effect transistor technology, *Nature Electronics*, **5**, 611–619 (2022).

17  F. Hakim, T. Tharpe, R. Tabrizian, Ferroelectric-on-Si Super-High-Frequency Fin Bulk Acoustic Resonators With $Hf_{0.5}Zr_{0.5}O_2$ Nanolaminated Transducers, *IEEE Microwave and Wireless Components Letters*, **31**, 701-704 (2021).

18  M. Ghatge, G. Walters, T. Nishida, R. Tabrizian, An ultrathin integrated nanoelectromechanical transducer based on hafnium zirconium oxide, *Nature Electronics*, **2**, 506-512 (2019).

19  S. S. Cheema, N. Shanker, LC. Wang, et al. Ultrathin ferroic $HfO_2$–$ZrO_2$ superlattice gate stack for advanced transistors, *Nature*, **604**, 65–71 (2022).

20  X. Zheng, T. Tharpe, S. M. E. H. Yousuf, N. G. Rudawski, P. X-L. Feng, R. Tabrizian, High Quality Factors in Superlattice Ferroelectric $Hf_{0.5}Zr_{0.5}O_2$ Nanoelectromechanical Resonators, *ACS Applied Materials & Interfaces,* **14**, 36807-36814 (2022).

21  S. Ghaffari, S. Chandorkar, S. Wang, E. J. Ng, C. H. Ahn, V. Hong, Y. Yang, T. W. Kenny, Quantum Limit of Quality Factor in Silicon Micro and Nano Mechanical Resonators, *Sci Rep*, **3**, 3244 (2013).

22  R. Tabrizian, M. Rais-Zadeh, F. Ayazi, Effect of phonon interactions on limiting the f.Q product of micromechanical resonators, *2009 International Solid-State Sensors, Actuators and Microsystems Conference*, **2009**, 2131-2134.

23  J.M.L. Miller, A. Ansari, D. B. Heinz, Y. Chen, I. B. Flader, D. D. Shin, L. G. Villanueva, T. W. Kenny, Effective quality factor tuning mechanisms in micromechanical resonators, *Appllied Physics Review*, **5**, 041307 (2018).

24  T. N. Tanskaya, V. N. Zima, A. G. Kozlov, The influence of surface roughness of Bragg reflector layers on characteristics of microwave solidly mounted resonator, *2015 International Siberian Conference on Control and Communications (SIBCON)*, **2015** 1-4.

25   A. Vorobiev, S. Gevorgian, M. Löffler, E. Olsson, Correlations between microstructure and Q-factor of tunable thin film bulk acoustic wave resonators, *Journal of Applied Physics*, **110**, 054102 (2011).

26   M. Ramezani, V. V. Felmetsger, N. G. Rudawski, R. Tabrizian, Growth of C-Axis Textured AlN Films on Vertical Sidewalls of Silicon Microfins, *Transactions on Ultrasonics, Ferroelectrics, and Frequency Control*, **68**, 753-759 (2021).

27   F. Martin, M. E. Jan, S. Rey-Mermet, B. Belgacem, D. Su, M. Cantoni, P. Muralt, Shear mode coupling and tilted grain growth of AlN thin films in BAW resonators, *IEEE Transactions on Ultrasonics, Ferroelectrics, and Frequency Control*, **53**, 1339-1343 (2006).

28   T. Tharpe, X. Q. Zheng, P. X. L. Feng, R. Tabrizian, Resolving Mechanical Properties and Morphology Evolution of Free-Standing Ferroelectric $Hf_{0.5}Zr_{0.5}O_2$, *Advanced Engineering Materials*, **23**, 2101221 (2021).

29   S. S. Fields, D. H. Olson, S. T. Jaszewski, C. M. Fancher, S. W. Smith, D. A. Dickie, G. Esteves, M. D. Henry, P. S. Davids, P. E. Hopkins, J. F.; Ihlefeld, Compositional and phase dependence of elastic modulus of crystalline and amorphous $Hf_{1-x}Zr_xO_2$ thin films, *Appl. Phys. Lett*. **118**, 102901 (2021).

30   C. Mart, T. Kämpfe, R. Hoffmann, S. Eßlinger, S. Kirbach, K. Kühnel, M. Czernohorsky, L. M. Eng, W. Weinreich, Piezoelectric Response of Polycrystalline Silicon-Doped Hafnium Oxide Thin Films Determined by Rapid Temperature Cycles, *Advanced Electronic Materials*, **6**, 1901015 (2020).

31   T. S. Mikolajick, S. Slesazeck, H. Mulaosmanovic, M. H. Park, S. Fichtner, P. Lomenzo, M. Hoffmann, U. Schroeder, Next generation ferroelectric materials for semiconductor process integration and their applications, *Journal of Applied Physics*, **129**, 100901 (2021).

32   O. J. Zobel, Theory and Design of Uniform and Composite Electric Wave Filters, *The Bell System Technical Journal*, **2**, 1-46 (1923).

33   M. Ghatge, R. Tabrizian, Dispersion-Engineered Guided-Wave Resonators in Anisotropic Single-Crystal Substrates—Part I: Concept and Analytical Design, *IEEE Transactions on Ultrasonics, Ferroelectrics, and Frequency Control*, **66**, 1140-1148 (2019).

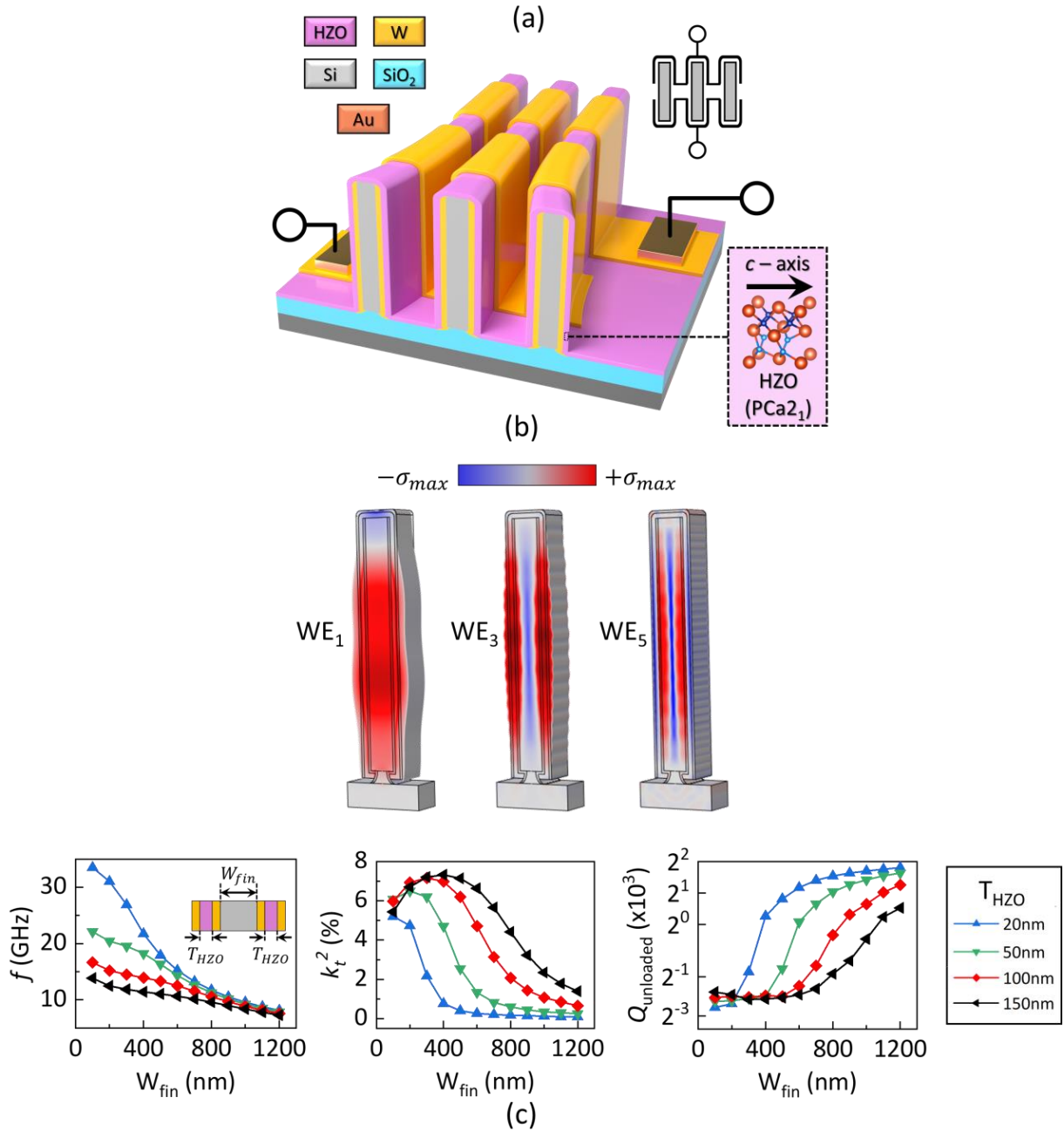

**Figure 1. Ferroelectric-gate fin (FGF) nano-acoustic resonator concept and scaling characteristics.** (a) Schematic demonstration of FGF resonator created from three parallel fins. The resonator relies on two three-dimensional gates that are electrically coupled through the ferroelectric HZO transducer and bottom floating W electrodes. The left inset shows proposed symbol for the one-port device. (b) COMSOL-simulated mode-shapes for FGF resonator operating in width-extensional bulk acoustic wave modes with odd harmonics of 1, 3, and 5. (c) Analytically modeled frequency, $Q$, and $k_t^2$ scaling characteristics of FGF resonators for different Si fin width and HZO transducer thicknesses.

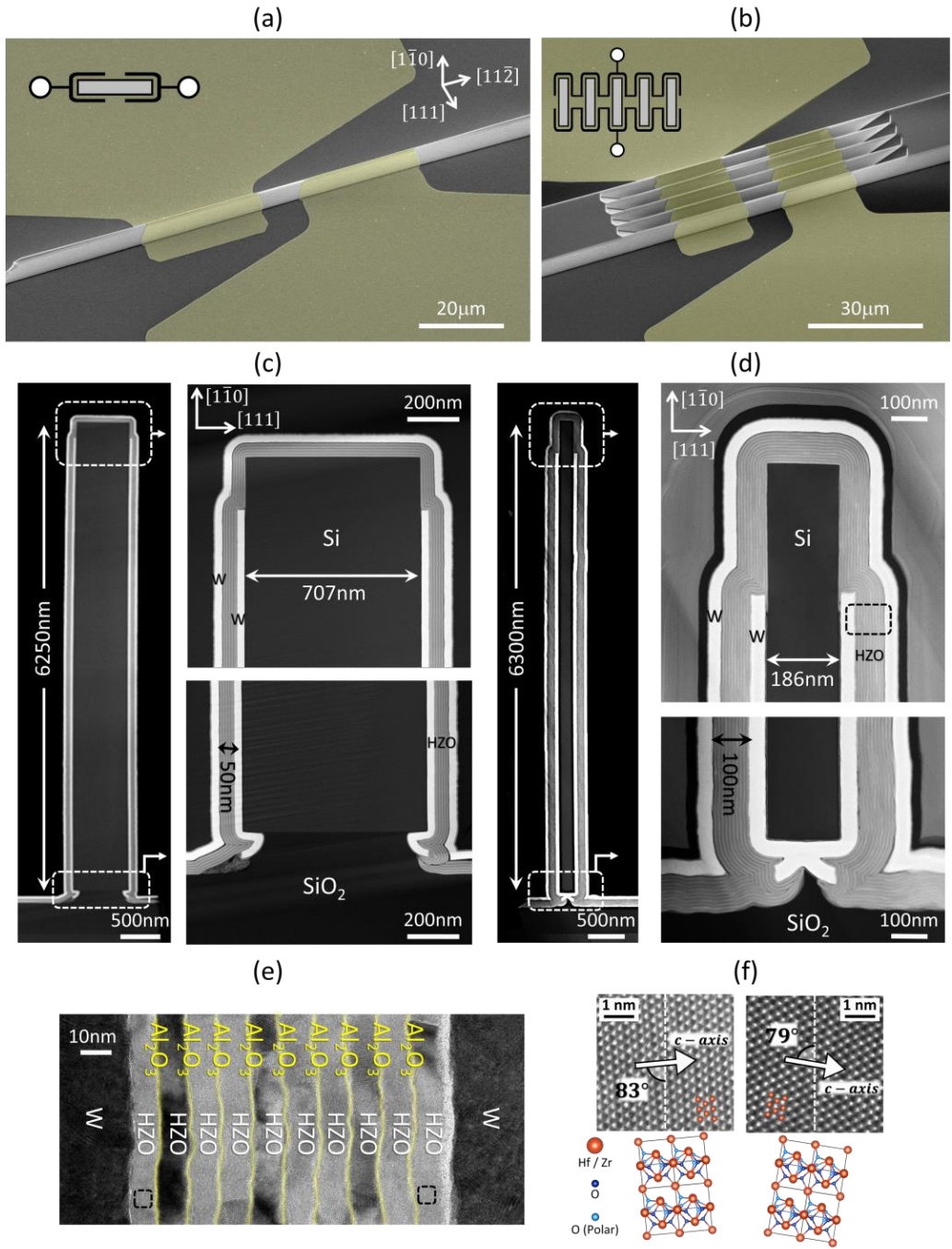

**Figure 2. Three-dimensional and cross-sectional images of ferroelectric-gate fin acoustic resonators.** SEM image of FGF resonator with (a) single fin and (b) five parallel fins. SEM and TEM image FGF resonators with (c) 707nm Si fin width and (d) 186nm Si fin width, and aspect ratio exceeding 34:1. Zoomed-in images at top and bottom corners of FGF resonators are also shown, highlighting three-dimensional patterning of W electrodes and conformal coverage of HZO transducer. (e) TEM image of HZO superlattice transducer, highlighting alumina interlayers that are used to sustain polar orthorhombic phase with large piezoelectricity. (f) HAADF-TEM images obtained from the selected areas in HZO transducer and their reconstructed atomic structure highlighting *c*-axis orientation.

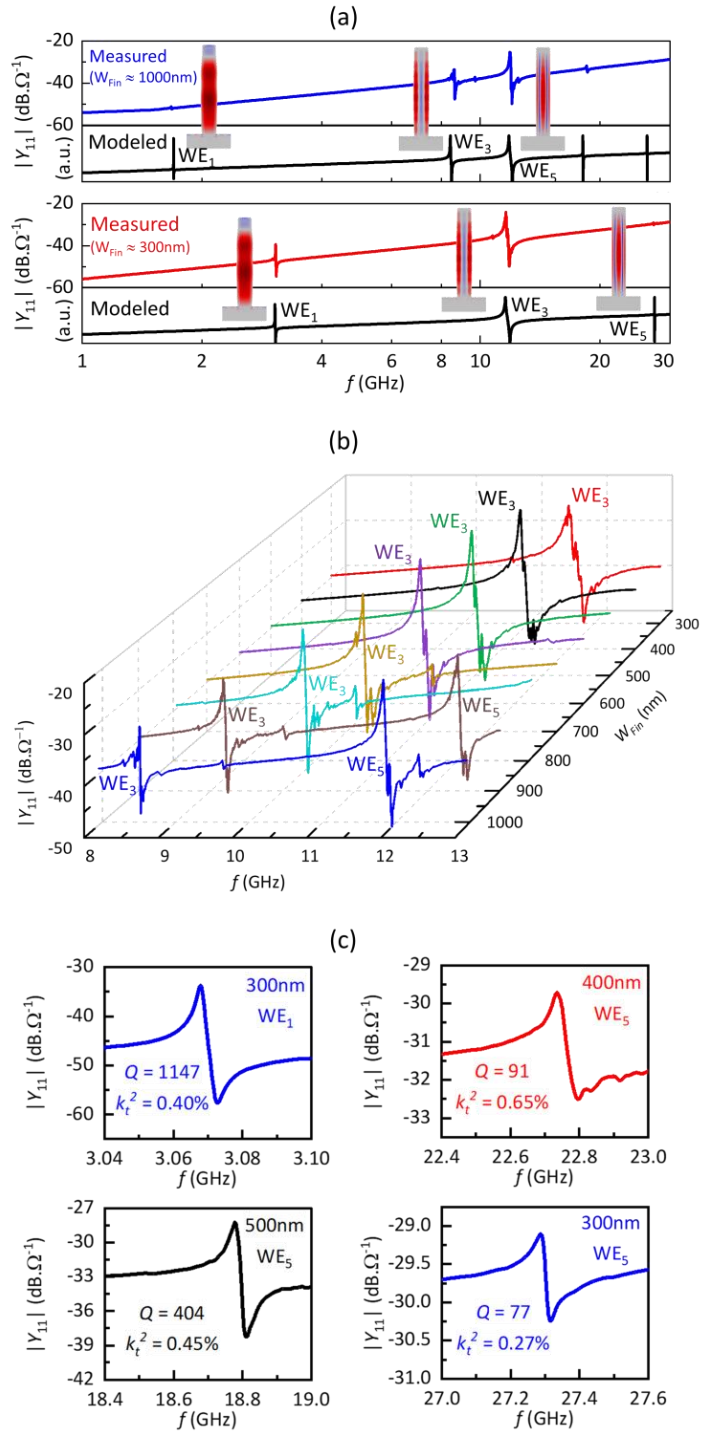

**Figure 3. Measured admittance of FGF nano-acoustic resonators.** (a) Large-span admittance for FGF resonators with fin width of 1000nm and 300nm, highlighting various width-extensional harmonics over 3-30 GHz. The measured admittances are compared with lossless Mason's model to identify the mode corresponding to each peak. (b) Admittance of resonators close to $WE_3$ mode, for fin widths over 300nm to 1000nm. (c) Short-span admittance for various peaks over 18-30 GHz. The fin width and WE modes are noted for each admittance plot.

-21-

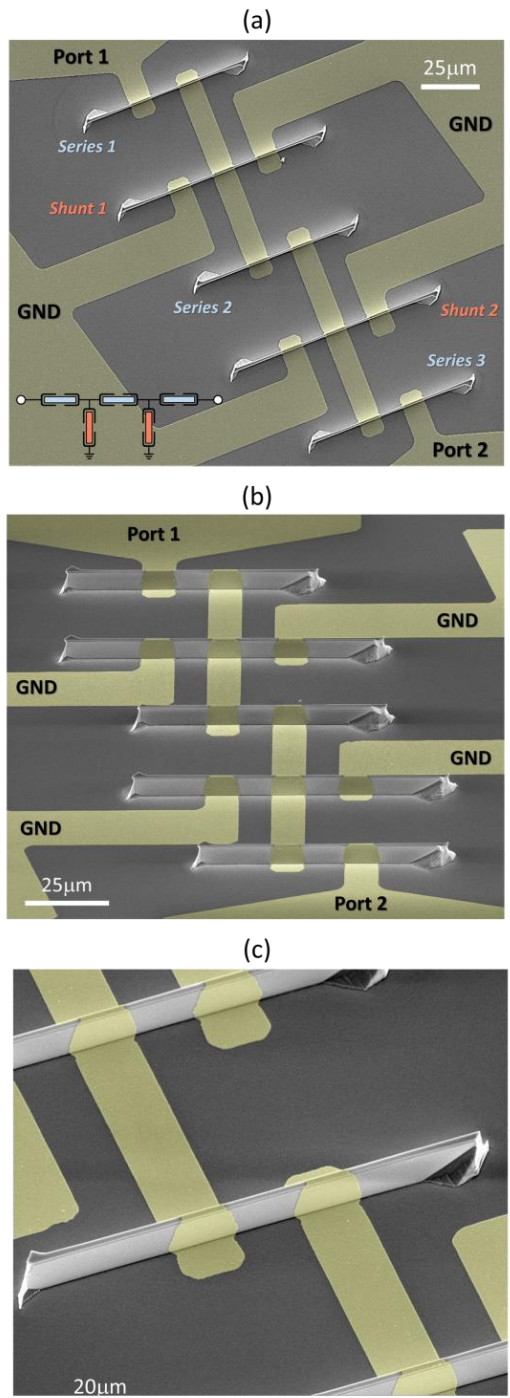

**Figure 4. Images of implemented FGF filter array.** (a) SEM image of FGF filter array to cover 9-12 GHz with lithographically tailorable center-frequencies and bandwidths. (b) Side-view of FGF filters created from five resonators connected in 2.5-stage ladder configuration. The series and shunt elements creating the ladder filter, highlighted in SEM image, are connected through planar routings. (c) Zoomed-in SEM image of filter around a three-dimensional gate on a series element, showing a slight taper across fin height. This is due to the slightly different exposure dose on piezoelectric mask across fin height.

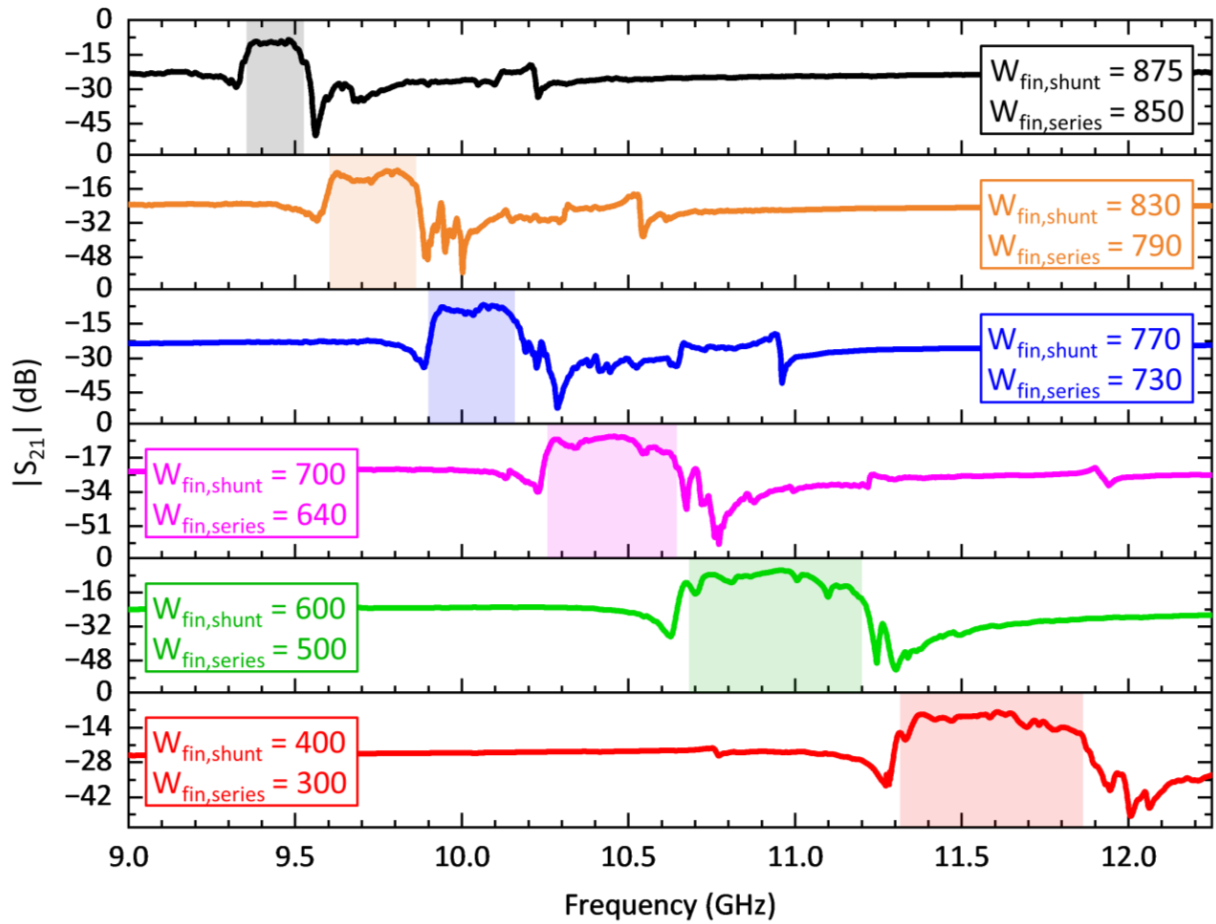

**Figure 5. Measured transmission response of FGF filter array.** $|S_{21}|$ of six filters created from FGF resonators with different fin widths, covering 9-12 GHz. Each filter is created from FGFs with slightly different fin widths to enable formation of highly-selective passband when resonators are coupled in ladder configuration. The approximate fin width of resonators for each filter is shown in legend.